\documentclass[onecolumn,aps,floatfix,superscriptaddress]{revtex4}
\usepackage{amsmath,amssymb,bm,bbm,eucal,graphicx}

\begin{document}

\title{Evolution of Cooperation by phenotypic similarity}
\author{Tibor Antal}
\affiliation{Program for Evolutionary Dynamics and Department of Mathematics, Harvard University, Cambridge MA 02138, USA}
\author{Hisashi Ohtsuki}
\affiliation{Department of Value and Decision Science, Tokyo Institute of Technology, Tokyo 152-8552, Japan}
\affiliation{PRESTO, Japan Science and Technology Agency, Saitama 332-0012, Japan}
\author{John Wakeley}
\affiliation{Department of Organismic and Evolutionary Biology, Harvard University, Cambridge MA 02138, USA}
\author{Peter D. Taylor}
\affiliation{Department of Mathematics and Statistics, Queen's University, Kingston, Ontario, Canada  K7L 3N6}
\author{Martin A. Nowak}
\affiliation{Program for Evolutionary Dynamics and Department of Mathematics, Harvard University, Cambridge MA 02138, USA}
\affiliation{Department of Organismic and Evolutionary Biology, Harvard University, Cambridge MA 02138, USA}

\begin{abstract}
The emergence of cooperation in populations of selfish individuals is a fascinating topic that has inspired much work in theoretical biology. Here we study the evolution of cooperation in a model where individuals are characterized by phenotypic properties that are visible to others. The population is well-mixed in the sense that everyone is equally likely to interact with everyone else, but the behavioral strategies can depend on distance in phenotype space. 
We study the interaction of cooperators and defectors. In our model, cooperators cooperate with those who are similar and defect otherwise. Defectors always defect. Individuals mutate to nearby phenotypes, which generates a random walk of the population in phenotype space. Our analysis bring together ideas from coalescence theory and evolutionary game dynamics. We obtain a precise condition for natural selection to favor cooperators over defectors. Cooperation is favored when the phenotypic mutation rate is large and the strategy mutation rate is small. In the optimal case for cooperators, in a one-dimensional phenotype space and for large population size, the critical benefit-to-cost ratio is given by $b/c=1+2/\sqrt{3}$. We also derive the fundamental condition for any two-strategy symmetric game and consider high-dimensional phenotype spaces.
\end{abstract}

\maketitle

\section{Introduction}

Evolutionary game theory is the study of frequency-dependent selection \cite{{msmith73},{taylor78},{msmith82},{hoffbauer98},{cressman03},{vincent05},{nowak04},{may73}}. Fitness values depend on the relative abundance, or frequency, of various strategies in the population, for example the frequency of cooperators and defectors. Evolutionary game theory has been applied to understand the evolution of cooperative interactions in viruses, bacteria, plants,  animals and humans \cite{{parker74},{colman95},{sinervo96},{nee00},{kerr02}}. The classical approach to evolutionary game dynamics assumes well-mixed populations, where every individual is equally likely to interact with every other individual \cite{{hoffbauer98}}. Recent advances include the extension to populations that are structured by geography or other factors \cite{{nowak92},{durrett94},{hassell94},{killingback96},{nakamaru97},{eshel99},{neu99},{szabo02},{hauert04},{ohtsuki06},{santos06},{taylor07}}. 

The term `greenbeard effect' was coined in sociobiology to describe the result of the following thought experiment  \cite{{hamilton64},{dawkins76}}.  What evolutionary dynamics will occur if a single gene is responsible for both a phenotypic signal (`a green beard') and a behavioral response (for example, altruistic behavior towards individuals with like phenotypes)? Later, the term `armpit effect' was introduced \cite{{dawkins82}} to refer to a self-referent phenotype that is used in identifying kin \cite{{mateo00},{sinervo06},{lize06}}.

Both of these concepts are now seen as cases of `tag-based cooperation', in which a generic system of phenotypic tags is used to indicate similarity or difference, and the evolutionary dynamics of cooperation are studied in the context of these tags. A first approach, based on computer simulations, assumed a well-mixed population, a continuum of tags, and an evolving threshold distance for cooperation \cite{riolo01}.  More recent models use numerical and analytic methods and often combine tags with viscous population structure \cite{{axelrod04},{jansen06},{hammond06},{rousset07},{gardner07}}.  A general finding of these papers is that it is difficult to obtain cooperation in  tag-based models for well-mixed populations, indicating that some spatial structure is needed \cite{nowak92}.  

Inspired by work on tag-based cooperation \cite{{riolo01},{hochberg03},{axelrod04},{jansen06}} and building on a previous approach \cite{{traulsen07}}, we study evolutionary game dynamics in a model where the behavior depends on phenotypic distance \cite{{levin82},{levin85}}. As a particular example we explore the evolution of cooperation \cite{{axelrod81},{nowak06}}. Studies of different organisms, including humans, support the idea that cooperation is more likely among similar individuals \cite{{lize06},{byrne69},{nahemow75},{selfhaut07}}. Our model applies to situations where individuals tend to like those who have similar attitudes and beliefs. We introduce a novel yet natural model in which individuals mutate to adjacent phenotypes in a possibly multi-dimensional phenotype space. We study one and infinitely many dimensions in detail. We develop a theory for general evolutionary games, not just the evolution of cooperation. Spatial structure is not needed for cooperation to be favored in our model.  Moreover, in contrast to previous work \cite{traulsen07}, we develop an analytic machinery for describing heterogeneous populations in phenotype space.

The paper is organized as follows. In Sec.~\ref{overview} we give an overview of the main results and provide a heuristic derivation. Then we derive the precise condition for cooperation to be favored. This condition depends on certain correlations in the neutral case, that is when each individual has the same fitness. These correlations are calculated in Section~\ref{neutral}, and in Section~\ref{critical} the condition for cooperation is derived. We relegate some details to the appendices. Finite population sizes are discussed in Appendix \ref{wf}, cooperation without self interaction in Appendix \ref{exclude}, and the derivation of correlations in Appendix \ref{avcorr}. In Appendix \ref{Moran} we show that all results in the large population size limit are identical for the W-F and for the Moran process. In Appendix \ref{general} we consider general payoff matrices, and finally in Appendix \ref{infdim} we discuss an infinite dimensional phenotype space.

\section{Overview of main results}
\label{overview}

Consider a population of asexual haploid individuals, with a population size $N$ that is constant over time. Each individual is characterized by a phenotype, given by an integer $i$ that can take any value from minus to plus infinity. Thus, this phenotype space is a one-dimensional and unbounded lattice. Individuals inherit the phenotype of their parent subject to some small variation. If the parent's phenotype is $i$, then the offspring has phenotype $i-1$, $i$ or $i+1$ with probabilities $v$, $1-2v$ and $v$, respectively. The parameter $v$ can vary between 0 and $1/2$.

\begin{figure}
\centering
\includegraphics[scale=0.5]{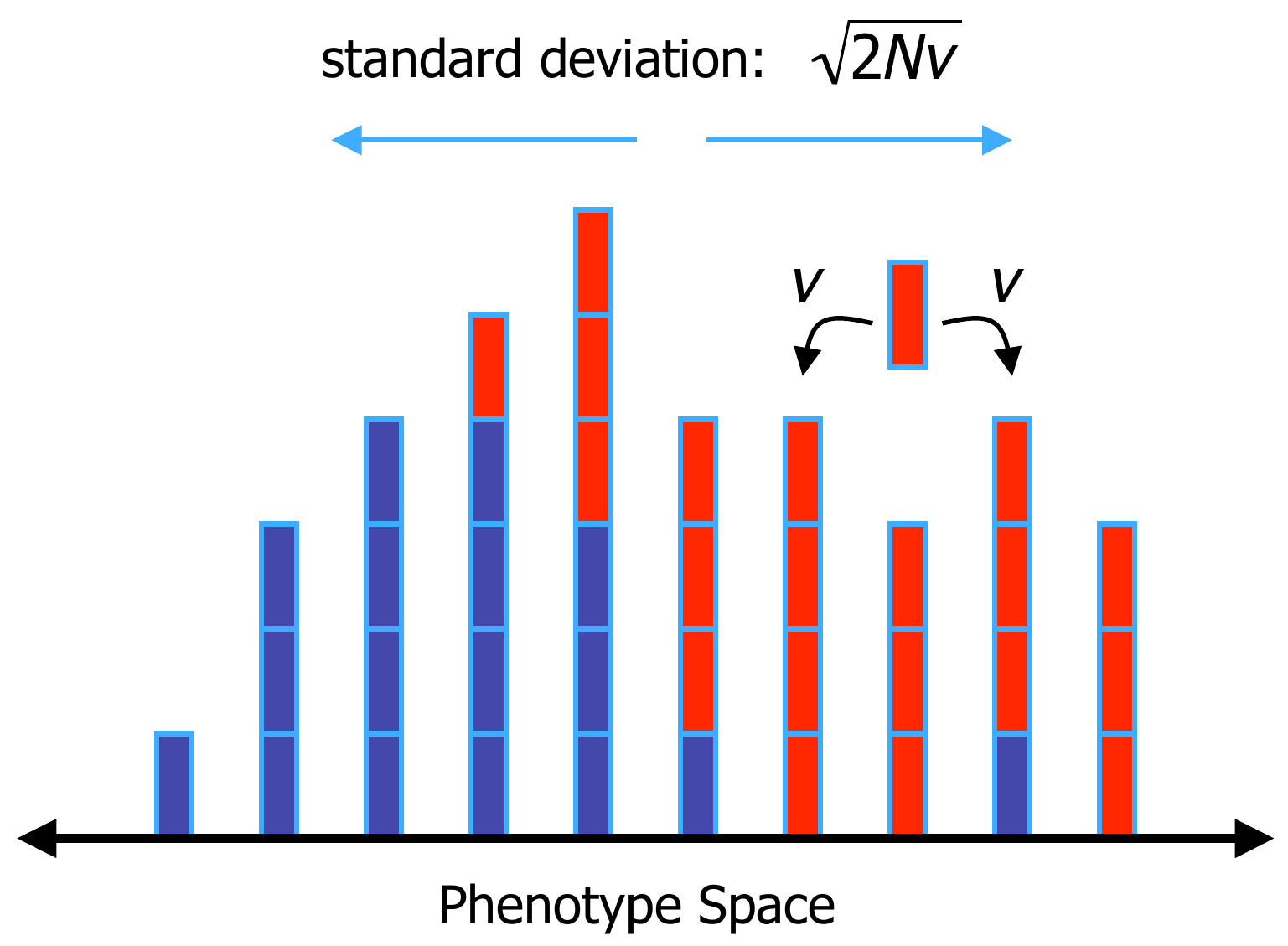}
\caption{The basic geometry of evolution in phenotype space. There are two types of individuals (red and blue), which can refer to arbitrary traits or different strategies in an evolutionary game. Individuals inherit the strategy of their parent subject to a small mutation rate $u$. Moreover, each individual has a phenotype. Here we consider a discrete one dimensional phenotype space. An individual of phenotype $i$ produces offspring of phenotype $i-1$, $i$ or $i+1$ with probabilities $v$, $1-2v$ and $v$, respectively. The total population (of size $N$) performs a random walk in phenotype space with diffusion coefficient $v$. Sometimes the cluster breaks into two or more pieces, but typically only one of them survives. If evolutionary updating occurs according to a Wright-Fisher process then the distribution of individuals in phenotype space has a standard deviation of $\sqrt{2Nv}$.  For the Moran process, the standard deviation is reduced to $\sqrt{Nv}$.}
\label{fig:model}
\end{figure}

Let us consider a Wright-Fisher process. In each generation, all individuals produce the same large number of offspring. The next generation of $N$ individuals is sampled from this pool of offspring.  To introduce some fundamental concepts and quantities, we first study the model without any selection.  No evolutionary game is yet being played, and there is only neutral drift in phenotype space.  The entire population performs a random walk with a diffusion coefficient $v$, and by this process will tend to disperse over the lattice.  In opposition to this, all of the individuals in the population will be, to some degree, related due to reproduction in a finite population. Thus, while occasionally the population may break up into two or more clusters, typically there is only a single cluster \cite{{moran75},{kingman76}}. The standard deviation of the distribution in phenotype space, which is a measure for the width of the cluster, is $\sqrt{2N v}$.  

\begin{figure}
\centering
\includegraphics[scale=0.3]{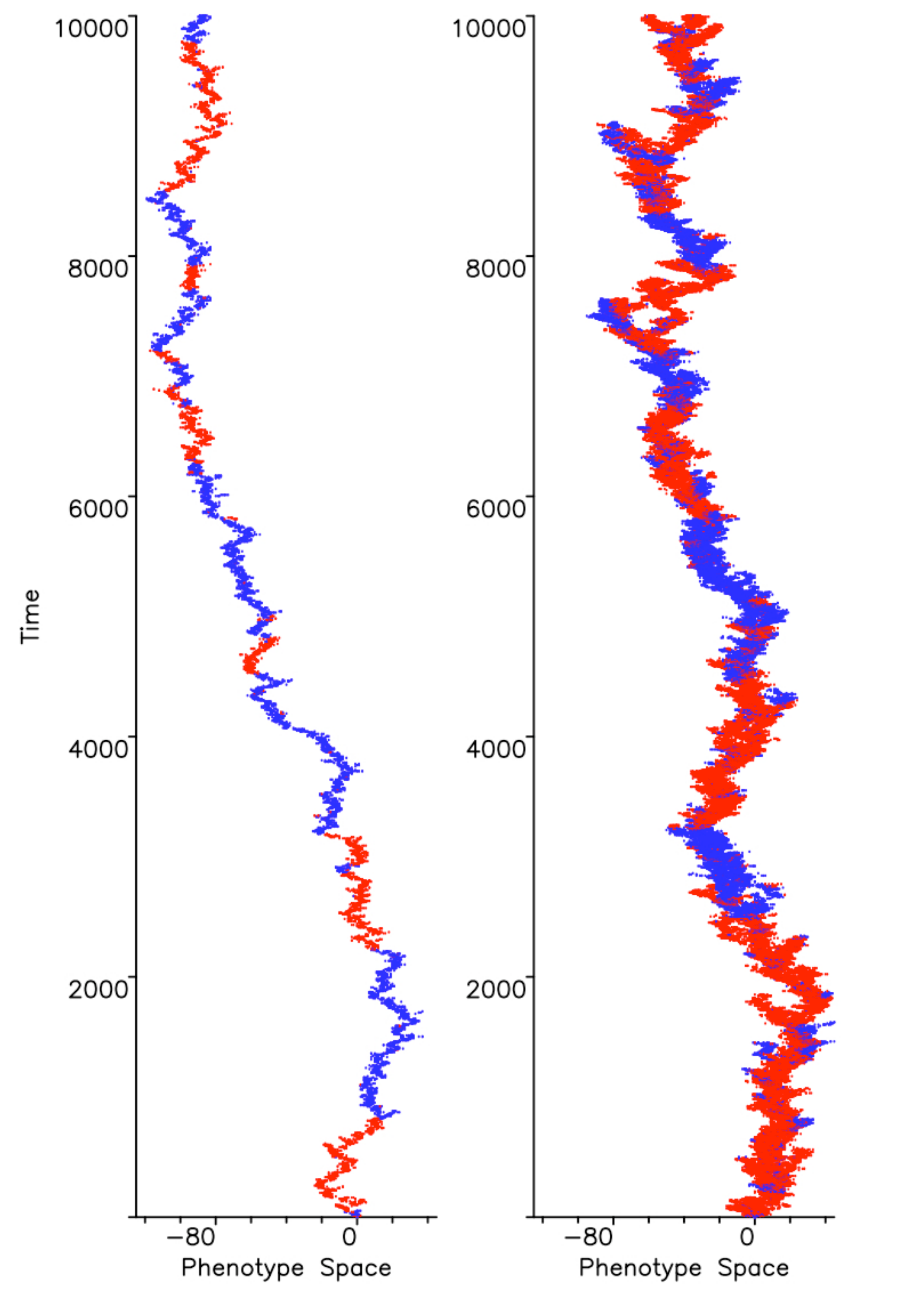}
\caption{Random walks in phenotype space. The figure shows two computer simulations of a Wright-Fisher process in a one-dimensional discrete phenotype space. The phenotypic mutation rate is $v=0.25$. The colors, red and blue, refer to arbitrary traits, because no game is yet being played. All individuals have the same fitness. 
The population size is (a) $N=10$ and (b) $N=100$. The strategy mutation probability (between red and blue) is $u=0.004$. Therefore, a given color dominates on average for $2/u=500$ generations (since new mutations arrive at rate $Nu/2$ and fixate with probability $1/N$). The standard deviation of the distribution in phenotype space is $\sqrt {2N v}$. About 95\% of all individuals are within 4 standard deviations. Often the population fragments into two or several pieces, but only one branch survives in the long run. We use the statistics of these neutral `phenotypic space walks' for calculating the fundamental conditions of evolutionary games in the limit of weak selection.}
\label{fig:walk}
\end{figure}

Next, we superimpose the neutral drift of two types: the strategies $A$ and $B$ (Fig.~\ref{fig:model}).  Still for the moment assuming no fitness differences, we have reproduction subject to mutation between $A$ and $B$.  Specifically, with probability $u$ the offspring adopts a random strategy. The mutation-reproduction process defines a stationary distribution \cite{wright31}.  If $u$ is very small relative to $N$, the population tends to be either all-$A$ or all-$B$. If $u$ is large, the population tends toward one half $A$ and one half $B$.  Figure \ref{fig:walk} illustrates the random walk in phenotype space of the population comprised of the two types $A$ and $B$.

Using coalescence theory \cite{{kingman82},{wakeley08}} many interesting and relevant properties of the distributions of both the strategies and phenotypic tags can be calculated.  For example, the probability that two randomly chosen individuals have the same phenotype is $z=1/\sqrt{8Nv}$. The probability that two randomly chosen individuals have the same strategy and the same phenotype is $g=z(1-Nu/2)$. The probability that two individuals have the same strategy and a third individual has the same phenotype as the second is $h=z[1-Nu(2+\sqrt{3})/4].$ These results hold for large population size $N$ and small mutation rate $u$; more precisely, we assume large $Nv$ and small $Nu$. The relevance of $z$, $g$ and $h$ will become clear below. The expressions for $z$, $g$ and $h$ are derived for general $Nv$ and $Nu$ in Section \ref{neutral}, where they appear as eqs \eqref{dfinal}, \eqref{gfinal} and  \eqref{hfinal}, respectively.  

We can now use these insights to study game dynamics.  We investigate the competition of cooperators, $C$, and defectors, $D$. Cooperators play a conditional strategy: they cooperate with all individuals who are close enough in phenotype space and defect otherwise. The notion of being close enough is modeled by a lattice structure. In particular, a cooperator with phenotype $i$ cooperates only with other individuals of phenotype $i$. Defectors, in contrast, play an unconditional strategy: they always defect. Cooperation means paying a cost, $c$, for the other individual to receive a benefit $b$.  The larger the total payoff of an individual, interacting equally with every member of the population, the larger the number of offspring it will produce on average.  We want to calculate the critical benefit-to-cost ratio, $b/c$, that allows the game in phenotype space to favor the evolution of cooperation.  

\begin{figure}
\centering
\includegraphics[scale=0.7]{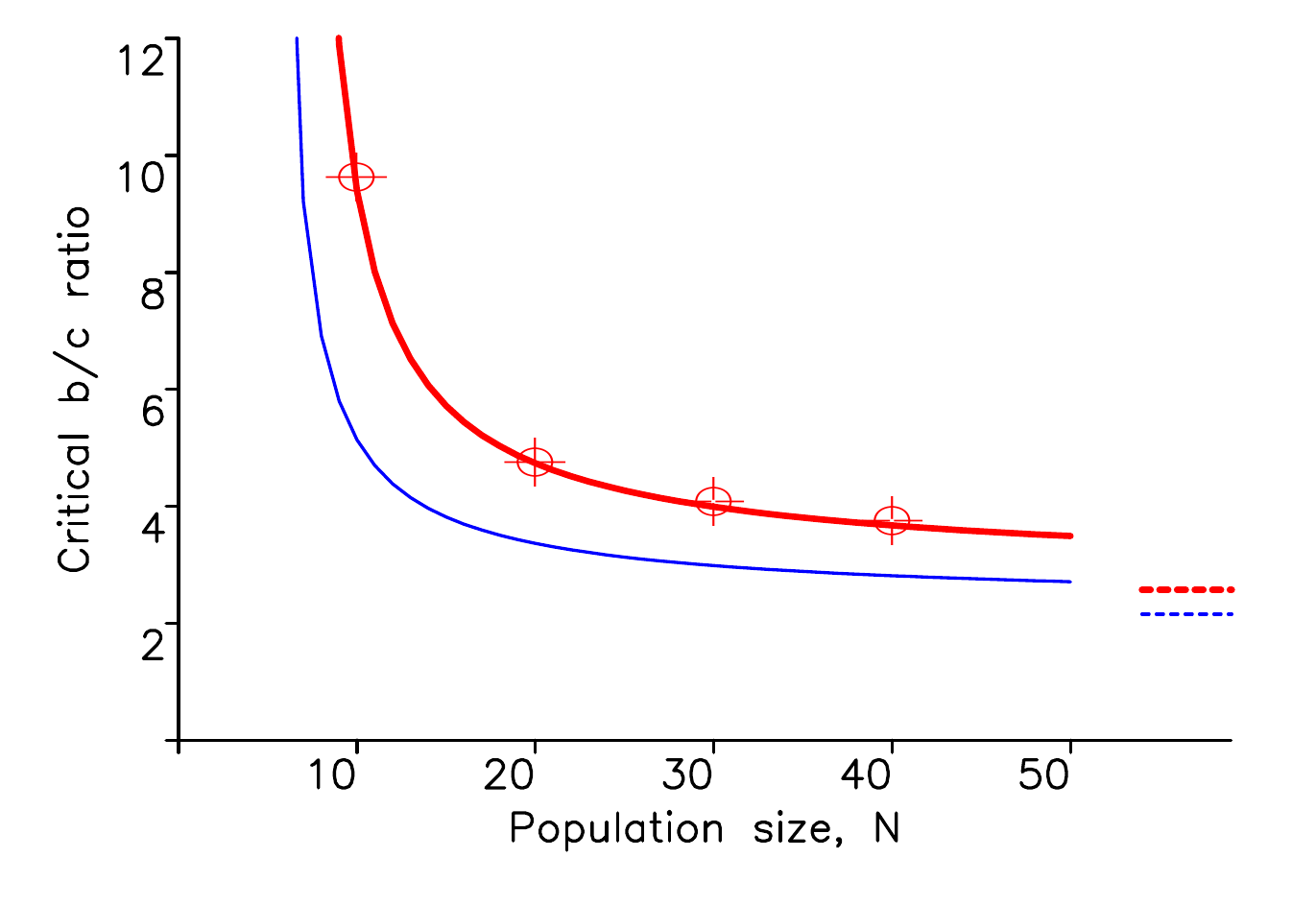}
\caption{Excellent agreement between numerical simulations and analytic calculations.  We show the critical benefit-to-cost ratio that is needed for cooperators to be more abundant than defectors in the stationary distribution. We have used a Wright-Fisher process with a phenotypic mutation rate $v=1/2$ and a strategy mutation probability $u=1/(2N)$. The red line indicates the result of our analytic calculation. For these parameter values the asymptotic limit for large $N$ is $b/c=(1+12\sqrt{2})/7\approx 2.5672$. The red dots indicate the result of numerical simulations. The grey line illustrates the critical $b/c$-ratio for $u\to 0$ with the asymptotic limit $b/c=1+2/\sqrt{3}\approx 2.1547$.}
\label{fig:fit}
\end{figure}

A configuration of the population is specified by $m_i$ and $n_i$, which are the number of cooperators  with phenotype $i$ and the total number of individuals with phenotype $i$, respectively. The total payoff of all cooperators is $F_C=\sum_i m_i (b m_i - c n_i)$. The total payoff of all defectors is $F_D=\sum_i (n_i-m_i) b m_i$. There are $\sum_i m_i$ cooperators and $N-\sum_i m_i$ defectors. The average payoff for a cooperator is $f_C=F_C/\sum_i m_i$. The average payoff for a defector is $f_D=F_D/(N-\sum_i m_i)$. Cooperators have a higher fitness than defectors if $f_C>f_D$, which leads to $\sum_im_i(bm_i-cn_i) > \sum_i m_i \sum_j m_jn_j (b-c)/N$. Averaging these quantities over every possible configuration of the population, weighted by their stationary probability under neutrality, we obtain the fundamental condition
\begin{equation}
\label{avercond}
b\Big< \sum_i m_i^2 \Big> - c\Big< \sum_im_in_i \Big> 
> (b-c)\Big< \sum_{ij} m_im_jn_j \Big>/N.
\end{equation}
Under this condition cooperators are more abundant than defectors in the mutation-selection process.
The above argument and our results are valid in the weak selection limit. A precise derivation of  this inequality is presented in Section \ref{critical}.  Correlation terms similar to the ones above sometimes arise in studies of social behavior and population dynamics \cite{{hamilton64},{price70}}. The first two terms in inequality \eqref{avercond} are pairwise correlations, while the third is notably a triplet correlation. Note that the argument leading to inequality \eqref{avercond} includes self-interaction, but that the effect of this becomes negligible when $N$ is large. 

When the population size is large, the averages in inequality \eqref{avercond} are proportional to the probabilities $g,z$ and $h$ respectively, which we introduced earlier.
Consequently, inequality \eqref{avercond} can be written as $bg-cz>(b-c)h$. Using the values of $z,g,h$ given above we obtain
\begin{equation}
\label{condi}
b/c > 1+ {\frac{2}{\sqrt 3}},
\end{equation}
which is approximately $2.16$.
If the benefit-to-cost ratio exceeds this number, then cooperators are more abundant than defectors in the mutation-selection process.  The success of cooperators results from the balance of movement and clustering in phenotype space.
Inequality \eqref{condi} represents the condition for cooperators to be more abundant than defectors in a large population when the strategy mutation rate $u$ is small ($Nu \ll 1$) and the phenotypic mutation rate $v$ is large ($Nv \gg 1$). We derive conditions for any population size and mutation rates. Figure \ref{fig:fit} shows the excellent agreement between numerical simulations and analytical calculations. In general we find that both lowering strategy mutations, and increasing phenotypic mutations favor cooperators.

\section{Correlations in the neutral case}
\label{neutral}

Let us give a more precise definition of the model first. Consider a population of $N$ haploid individuals (players). Each individual $k=1,\dots,N$ has an integer-valued phenotype $X_k\in \mathbb{Z}$, which we also refer to as its position in phenotype space. Additionally, each individual has a strategy $S_k\in\{0,1\}$, and we refer to these two strategies as cooperation (1) and defection (0). In general, players' phenotypes and strategies determine their fitness. In the Wright-Fisher (W-F) process, each of the $N$ individuals of the next generation independently chooses a parent from the previous generation with a probability proportional to the parent's fitness. Each offspring inherits the parent's position (phenotype) with probability $1-2v$, and it is placed to either the left or the right neighboring position of the parent, both with probability $v$. Each offspring also inherits the parent's strategy with probability $1-u$, and it adopts a random strategy with probability $u$.

In this section we consider the neutral case, that is when all players have the same fitness.
Note that the strategies and the phenotypes of the individuals change independently, and evolve according to the Wright-Fisher process \cite{{moran75},{kingman76}}. The system rapidly reaches a stationary state where the individuals stay in a cluster with variance $2Nv$, but the cluster as a whole diffuses over the space (the integers) with diffusion coefficient $v$. We are interested in the properties of this stationary state.

We are particularly interested in four probabilities. We pick three distinct individuals $k$, $q$, and $l$ from the population in the stationary state. For their phenotypes and their strategies we define the following four probabilities
\begin{equation}
\label{corrdefs}
\begin{split}
  y &= \mathrm{Pr} (S_k=S_q)\\
  z &= \mathrm{Pr} (X_k=X_q)\\
  g  &= \mathrm{Pr} (S_k=S_q,~ X_k=X_q)\\
  h &= \mathrm{Pr} (S_l=S_k,~ X_k=X_q )
\end{split} 
\end{equation}
In words, $y$ is the probability that two individuals have the same strategy, and $z$ is the probability that they have the same phenotype. They have simultaneously the same strategy and phenotype with probability $g$. Out of three individuals, the probability that the first two have the same phenotype, and simultaneously the first and the third have the same strategy is denoted by $h$. Note that neither $g$ nor $h$ factorizes in general.

To obtain the above probabilities we have to know the probability $\mathrm{Pr}(T=t)$ that the time $T$ to the most recent common ancestor (MRCA) of two randomly chosen individual is $T=t$. This time is not affected by either the strategies or the phenotypes of the players. It is determined solely by the W-F dynamics. The ancestry of two individuals coalesce with probability $1/N$ in each time step. Hence the probability that the time to the MRCA is $t$ is
\begin{equation}
\label{timewf}
 \mathrm{Pr}(T=t) = \left( 1-\frac{1}{N} \right)^{t-1} \frac{1}{N}
\end{equation}

We can continue the calculation for finite system size $N$, but the expressions become cumbersome. Hence we relegated the finite $N$ calculations to Appendix \ref{wf}, where we mainly treat the special $v=1/2$ case. In this section we discuss the large population limit $N\to\infty$, where we introduce the rescaled time $\tau=t/N$. In this limit we can use a continuous time description, where the coalescent time distribution \eqref{timewf} is given by the density function
\begin{equation}
\label{coaltime}
 p(\tau) = e^{-\tau}
\end{equation}
and the average coalescence time becomes $\tau=1$ in the new unit.

Due to the non-overlapping generations in the W-F model, each individual is a newborn and has the chance to mutate both in strategy and phenotype space. In the large $N$ and $u,v\to0$ limit, the system can be described as a continuous time process. Strategy mutations arrive at rate $\mu=2Nu$ and phenotype mutations at rate $\nu=2Nv$ (in each direction) on the ancestral line of {\it two} individuals. Note that this continuous time limit is exact for the Moran process even for finite values of $v$, as it is shown in Appendix \ref{Moran}. In the W-F model, for finite values of $v$ we have a discrete time random walk, but the typical number of steps goes to infinity. In that limit the discrete and continuous time walks become identical, and hence the finite $v$ behavior can be recovered as the $\nu\to\infty$ limit.

\subsection{Phenotypic distance}

Let us first study the phenotypes of the players. Here we calculate not only $z$, but in general the probability that two randomly chosen individuals $k$ and $q$ are at distance $x$ in phenotype space
\begin{equation}
\label{defdis}
 z(x) = \mathrm{Pr}(X_k-X_q=x)
\end{equation}
We know that the (signed) distance between the two individuals changes by plus or minus one at rate $\nu$, and the distance distribution after time $\tau$ can be expressed in terms of the Modified Bessel functions \cite{kampen97,redner01} as
\begin{equation}
\label{rw}
 \zeta(x| \tau) = e^{-2\nu\tau} I_{|x|}(2\nu\tau)
\end{equation}
The probability that two individuals are distance $x$ apart is
\begin{equation}
\label{dismethod}
 z(x) = \sum_{t=1}^\infty \mathrm{Pr}(X_k-X_q=x | T=t) \mathrm{Pr}(T=t)
\end{equation}
which becomes an integral of the corresponding density functions in the continuous time limit
\begin{equation}
 z(x) = \int\limits_0^\infty p(\tau) \zeta(x|\tau) d\tau = \int\limits_0^\infty e^{-(2\nu+1)\tau} I_{|x|}(2\nu\tau) d\tau
\end{equation}
By using the identity \cite{gradshteyn}
\begin{equation}
\label{bessel}
 \int\limits_0^\infty e^{-ac}\, I_\gamma (bc)\, dc = \frac{b^{-\gamma}\left( a-\sqrt{a^2-b^2}\right)^\gamma}{\sqrt{a^2-b^2}}
\end{equation}
we arrive at the probability distribution of the signed distance
\begin{equation}
\label{distance}
 z(x) = \frac{1}{\sqrt{4\nu+1}} \left( \frac{2\nu+1-\sqrt{4\nu+1}}{2\nu} \right)^{|x|}
\end{equation}
The individuals are at the same position with probability
\begin{equation}
\label{dfinal}
 z\equiv z(0) = \frac{1}{\sqrt{4\nu+1}} 
\end{equation}
Distribution \eqref{distance} is of course normalized $\sum_{x=-\infty}^\infty z(x)=1$, and its second moment is 
\begin{equation}
\sum_{x=-\infty}^\infty x^2 z(x)=2\nu
\end{equation}
Note that this second moment is twice the variance of the individual positions, which is exactly $\nu=2Nv$ even for finite $N$ (see Appendix \ref{wf}). Hence the individuals stay together in a cluster of size $\sqrt{2Nv}$. This cluster diffuses collectively through phenotype space. If one follows the ancestral line of an individual time $\tau$ back, its position $\hat x(\tau)$ will change by one at rate $\nu/2$ in each direction. Consequently, the position of the cluster has a variance proportional to time
\begin{equation}
 \langle \hat x^2 \rangle = \nu\tau = 2v t
\end{equation}
which implies a diffusive motion. The same result is valid for any finite $N$ in the large time limit. Note that the diffusion coefficient $D=v$ does not depend on the population size. Since the cluster itself wanders in space, the average number of individuals at any given site goes to zero. That is why we focus on distances in the phenotype space \eqref{defdis}.

\subsection{Pair with same strategy}

We are interested in the probability $y$ that two randomly chosen individuals have the same strategy.
In the continuous time limit, strategy mutations arrive at rate $\mu$ on the ancestral lines of the two individuals. The two individuals have the same strategy if there were no mutations, which is the case with probability $e^{-\mu\tau}$. Otherwise there was at least one mutation, hence at least one of the players has a random strategy, so they have the same strategy with probability $1/2$. Consequently, the probability that two players have the same strategy time $\tau$ after their MRCA is
\begin{equation}
\label{even}
 y(\tau) =  e^{-\mu\tau} + \frac{1}{2} \left(1-e^{-\mu\tau}\right)
\end{equation}
The probability $y$ that two randomly chosen individuals have the same strategy is
\begin{equation}
 y = \sum_{t=1}^\infty \mathrm{Pr}(S_k=S_q| T=t) \mathrm{Pr}(T=t)
\end{equation}
In the continuous time limit we obtain
\begin{equation}
\label{sigma}
 y = \int\limits_0^\infty p(\tau) y(\tau) d\tau = \frac{2+\mu}{2(1+\mu)}
\end{equation}
where we have used \eqref{coaltime} and \eqref{even}. 

\subsection{Pair with same strategy and phenotype}

The probability $g$ that two randomly chosen individuals have the same phenotype and also have the same strategy can be obtained as
\begin{equation}
\label{gmethod}
 g = \sum_{t=1}^\infty \mathrm{Pr}(S_k=S_q| T=t)  \mathrm{Pr}( X_k=X_q | T=t)  \mathrm{Pr}(T=t)
\end{equation}
Here we have used the property, that although $g$ does not factorize in general, nevertheless for any given time $t$ the conditional probabilities factorize as
\begin{equation}
 \mathrm{Pr}(S_k=S_q,~ X_k=X_q | T=t) =  \mathrm{Pr}(S_k=S_q| T=t)  \mathrm{Pr}( X_k=X_q | T=t) 
\end{equation}
The reason is that mutations occur completely independently in the strategy and the phenotype space. The corresponding integral in the continuous time limit hence becomes
\begin{equation}
\label{ccdef}
 g = \int\limits_0^\infty p(\tau) y(\tau) \zeta(\tau) d\tau 
\end{equation}
where we use the notation $\zeta(\tau)\equiv\zeta(0|\tau)$.  Note that it is also easy to obtain the analog probability where the phenotype difference is $x$, but we do not consider that here. Using identity \eqref{bessel} again, we can evaluate the above integral
\begin{equation}
\label{gfinal}
 g = \frac{1}{2\sqrt{1+4\nu}} + \frac{1}{2\sqrt{(1+\mu)(1+\mu+4\nu)}}
\end{equation}

\subsection{Three point correlations}

Now we turn to the calculation of the three point probability $h$ which is defined in \eqref{corrdefs}. If we follow the ancestral lines of three individuals back in time, the probability that there was no coalescence event  during one update step is $(1-1/N)(1-2/N)$. Two individuals coalesce with probability $3/N\cdot(1-1/N)$. When two individual have coalesced, the remaining two coalesce with probability $1/N$ during each update step. Hence the probability that the first merging happens to any pair of individuals at time $t_3\ge 1$ back in time, and the second $t_2\ge 1$ before the first one is
\begin{equation}
\label{tt1}
 \mathrm{Pr}(t_3, t_2) = \frac{3}{N^2} \left[ \left(1-\frac{1}{N}\right) \left(1-\frac{2}{N}\right)\right]^{t_3-1} \left(1-\frac{1}{N}\right)^{t_2} 
\end{equation}
The probability that three individual coalesce simultaneously at time $t_3$ is
\begin{equation}
\label{tt2}
 \mathrm{Pr}(t_3, 0) = \frac{1}{N^2} \left[ \left(1-\frac{1}{N}\right) \left(1-\frac{2}{N}\right)\right]^{t_3-1}
\end{equation}
In the $N\to\infty$ limit \eqref{tt1} converges to the density function
\begin{equation}
\label{threetime}
 p(\tau_3, \tau_2) = 3 e^{-(3\tau_3+\tau_2)}
\end{equation}
with $\tau_3=t_3/N$ and $\tau_2=t_2/N$. Note that \eqref{tt2} does not affect the large $N$ limit.

Let us call the scaled time when individuals $q,k$ coalesce $\tau_{qk}$, and when $k,l$ coalesce $\tau_{kl}$. With probability $1/3$ individuals $q,k$ coalesce first at $\tau_{qk}=\tau_3$ and they coalesce with $l$ at $\tau_{kl}=\tau_3+\tau_2$. Similarly with probability $1/3$ individuals $k,l$ coalesce first at $\tau_{kl}=\tau_3$ and they coalesce with $q$ at $\tau_{qk}=\tau_3+\tau_2$. If, however, $l,q$ coalesce first with probability 1/3, it makes $\tau_{qk}=\tau_{kl}=\tau_3+\tau_2$. Since we know the probability density $y(\tau)$ that two individuals with a MRCA at time $\tau$ back have the same strategy \eqref{even}, and the probability density $\zeta(\tau)\equiv\zeta(0|\tau)$ that they are at the same position \eqref{rw}, we can simply obtain the three point correlation as
\begin{equation}
\label{threeform}
 h = \frac{1}{3} \int\limits_0^\infty d\tau_3 \int\limits_0^\infty d\tau_2~~ p(\tau_3, \tau_2) 
  \left[ \zeta(\tau_3)y(\tau_3+\tau_2) 
 + \zeta(\tau_3+\tau_2)y(\tau_3) 
  + \zeta(\tau_3+\tau_2)y(\tau_3+\tau_2) \right]
\end{equation}
This integral can be evaluated by first introducing a variable for $\tau_2+\tau_3$ in the last two terms of the integral, and by using identity \eqref{bessel} in all three terms. We obtain
\begin{equation}
\label{hfinal}
 h = \frac{(1+\mu)(3+\mu)+ C_1 (2+\mu) - \mu C_3}{2(1+\mu)(2+\mu)\sqrt{1+4\nu}}
\end{equation}
with the shorthand notation
\begin{equation}
\label{shorty}
 C_i = \frac{1}{2}\sqrt\frac{(i+\mu)(1+4\nu)}{i+\mu+4\nu}
\end{equation}
By now we have obtained all the correlations in \eqref{corrdefs} in the $N\to\infty$ limit for any values of $\nu$ and $\mu$.


\section{Threshold $b/c$ ratio}
\label{critical}

In this section the individuals play a simplified Prisoner's Dilemma game given by the payoff matrix
\begin{equation}
\label{payoffmat}
\begin{tabular}{lc|cc}
 & & \multicolumn{2}{c} {when playing against}\\
 & & $C$ & $D$ \\ \hline
 & & & \\
 & $C$ & $~~~b-c$ & $-c$ \\ 
 payoff of & & &\\
 & $D$ & $b$ & $0$ \\ 
\end{tabular}
\end{equation} 
Here $b>0$ is the benefit gained from cooperators, and $c>0$ is the cost payed by cooperators.  We assume that all individuals interact (in this sense the population is ``well mixed"). Cooperators, however, play a conditional strategy: they cooperate with other individuals who have the same phenotype, and they defect otherwise. Defectors always defect. The total payoff of an individual is the sum of all payoffs that individual receives. We introduce the effective payoff of an individual $f=1+\delta \cdot \mathrm{payoff}$, where $\delta>0$ is the strength of the selection, and $\delta=0$ corresponds to the neutral case discussed in Section~\ref{neutral}. Note that $\delta$ must be sufficiently small to make all fitness values positive.

We consider here the simplest possible case, where each individual also receives a payoff from self interaction. Excluding self-interaction results in a $1/N$ correction, which is discussed in Appendix \ref{exclude}. An extension to a general payoff matrix is considered in Appendix \ref{general}. 


\subsection{Fitness}

Let $n_{i}$ denote the number of players of phenotype $i$, and $m_{i}$ the number of {\it cooperators} of phenotype $i$. A state of the system is given by the vectors $s=(\boldsymbol{n}, \boldsymbol{m})$. Let $f_{C,i}$ and $f_{D,i}$ represent the (effective) payoffs of a cooperator and a defector, respectively, of phenotype $i$. When self interaction is included these values are
\begin{equation}
\label{payoff}
	\begin{split}
		f_{C, i} &= 1+ \delta \left[ b m_{i} - c n_{i} \right]\\
		f_{D, i} &= 1+ \delta \left[ b m_{i} \right].
	\end{split}
\end{equation}

Let $w_{C,i}$ and $w_{D,i}$ represent the fitness ({\it i.e.} average number of offsprings) of a cooperator and a defector of phenotype $i$. After one update step (which is one generation) we obtain
\begin{equation}
\label{fittWF}
	\begin{split}
		w_{C, i} &= \frac{N f_{C,i}}{\sum_{j} [ m_{j} f_{C,j} + (n_{j}-m_{j}) f_{D,j}]}\\
	\end{split}
\end{equation}
Here a cooperator is chosen to be a parent with probability given by its payoff relative to the total payoff, and this happens $N$ times independently in one update step. The denominator of \eqref{fittWF} can be written as
\begin{equation}
	\sum_{j} [ m_{j} f_{C,j} + (n_{j}-m_{j}) f_{D,j}] = N + \delta (b-c) \sum_{j} m_{j} n_{j} 
\end{equation}
Therefore, in the $\delta\to 0$ limit, we obtain the fitness of a phenotype $i$ cooperator 
\begin{equation}
\label{fittfinalWF}
		w_{C, i} = 1 + \delta \left( b m_{i} - c n_{i} - \frac{b-c}{N} \sum_{j} m_{j} n_{j} \right) +  {\mathcal O} (\delta^{2})
\end{equation}

\subsection{Effect of selection}

Let $p$ denote the frequency of cooperators in the population. 
Cooperation is favored if cooperators are in the majority at the stationary state, $\langle p \rangle>1/2$.
The frequency of cooperators $p$ changes during one update step due to selection and due to mutation.
In any state $s$ of the system, the total change of cooperator frequency can be expressed in terms of the change due to selection as
\begin{equation}
\label{totchange}
 \Delta p_\mathrm{tot}(s) = (1-u) \Delta p_\mathrm{sel}(s) + u\left( \frac{1}{2}-p \right)
\end{equation}
Here the first term describes the change due to selection in the absence of mutation, which happens with probability $1-u$. The second term stands for the effect of mutation, which happens with probability $u$ to each player independently. In this latter case the frequency $p$ increases in average by $1/2$ due to the introduction of random strategies, and decreases by $p$ due to the replacement of cooperators. 

In the stationary state $\langle p \rangle$ is constant, hence the total change of frequency vanishes $\langle \Delta p\rangle_\mathrm{tot}=0$. Then from \eqref{totchange} we can express
the average cooperator frequency with the change of frequency due to selection as
\begin{equation}
\label{averwithsel}
 \langle p\rangle = \frac{1}{2} + \frac{1-u}{u} \langle \Delta p \rangle_\mathrm{sel}
\end{equation}
This means that by calculating the average change of cooperator frequency, we also obtain the average cooperator frequency. It also means that cooperators are favored $\langle p \rangle>1/2$ if their change due to selection is positive in the stationary state
\begin{equation}
\label{positivechange}
 \langle \Delta p \rangle_\mathrm{sel} > 0
\end{equation}

Now let us perform a perturbative expansion for small selection $\delta\ll 1$.
In a given state $s=(\boldsymbol{n}, \boldsymbol{m})$, the expected change of $p$ due to selection in one update step is
\begin{equation}
	\Delta p(s) = \frac{1}{N} \left( \sum_{i} m_{i} w_{C,i} - \sum_{i} m_{i} \right)
\end{equation}
This expression vanishes for $\delta=0$ for the fitness function \eqref{fittfinalWF}. (Note that this statement is not true in general for arbitrary models). Its Taylor expansion is
\begin{equation}
\label{deltap_s}
  \Delta p(s) = 0 + \delta \frac{d \Delta p(s)}{d \delta} \Big|_{\delta=0} +  {\mathcal O} (\delta^{2})
= \frac{\delta}{N} \sum_{i} m_{i} \frac{d w_{C,i}}{d \delta}\Big|_{\delta=0} +  {\mathcal O} (\delta^{2})
\end{equation}
We also expand the stationary probabilities of finding the system in state $s$
\begin{equation}
\label{statprob}
 \pi(s) = \pi^{(0)}(s) + \delta  \pi^{(1)}(s) + {\mathcal O} (\delta^{2})
\end{equation}
where $\pi^{(0)}(s)$ is the stationary probability in the neutral state
(here we consider two states equivalent if they only differ by translation along the phenotype space).
Consequently, in the stationary state in the presence of the game, the average change in cooperator frequency can be expressed in the leading order in terms of averages in the neutral stationary state
\begin{equation}
\label{dpsel0}
	\langle \Delta p \rangle_\mathrm{sel} = \frac{\delta}{N} \left\langle \sum_{i} m_{i} \frac{d w_{C,i}}{d \delta} \right\rangle_0 +  {\mathcal O} (\delta^{2})
\end{equation}
This expression has to be positive for cooperation to be favored \eqref{positivechange}. Here the 0  subscript refers to $\delta=0$, that is to an average taken in the stationary state of the neutral model $\langle\cdot\rangle_0 = \sum_s \cdot\, \pi^{(0)}(s)$. More generally, one can also easily obtain higher order terms in $\delta$ based on \eqref{deltap_s} and \eqref{statprob}. The first derivative of the effect of selection in the stationary state
\begin{equation}
 \left\langle \Delta p \right\rangle_\mathrm{sel}^{(1)} = \frac{d\langle \Delta p \rangle_\mathrm{sel}}{d\delta}\Big|_{\delta=0} 
\end{equation}
can be obtained from \eqref{dpsel0}, by using the fitness \eqref{fittfinalWF} of our model, as
\begin{equation}
\label{dpsel1}
	\left\langle \Delta p \right\rangle_\mathrm{sel}^{(1)} = \frac{1}{N} \left[ b \left\langle \sum_{i} m_{i}^{2} \right\rangle_0 - c  \left\langle \sum_{i} m_{i} n_{i} \right\rangle_0 - \frac{b-c}{N} \left\langle \sum_{i,j} m_{i} m_{j} n_{j} \right\rangle_0 \right]
\end{equation}
The threshold model parameters are then obtained when the change $\left\langle \Delta p \right\rangle_\mathrm{sel}^{(1)}=0$, as follows from the general condition \eqref{positivechange}
\begin{equation}
\label{critgenexp}
	\left(\frac{b}{c} \right)^* = \frac{\left\langle \sum_i m_i n_i \right\rangle_0-
	\frac{1}{N} \left\langle \sum_{i,j} m_i m_j n_j \right\rangle_0}
	{\left\langle \sum_i m_i^2 \right\rangle_0-
	\frac{1}{N} \left\langle \sum_{i,j} m_i m_j n_j \right\rangle_0}
\end{equation}
Hence, we have expressed the threshold $b/c$ ratio in the small selection limit in terms of correlations in the neutral stationary state. Note that the averages in \eqref{dpsel1} cannot be moved inside the sum, since at any given position any stationary average is zero. Also note that all terms in \eqref{critgenexp} are of order  $N^2$.

The above derivation is valid for finite $N$ and $\delta\to0$. We are also interested, however, in the $N\to\infty$ asymptotic behavior. In that case all the above derivation can be repeated when simultaneously $\delta N\to0$. 

Expression \eqref{dpsel1} for the change in cooperator frequency can be rewritten in a more intuitive way. First we express the total payoffs of cooperators and defectors respectively as
\begin{equation}
\begin{split}
 f_C &= \sum_i m_i f_{C,i} = N_C + \delta F_C\\
 f_D &= \sum_i m_i f_{D,i} = N_D + \delta F_D
\end{split}
\end{equation}
in a given state, where $F_C$ and $F_D$ are the total payoffs without considering weak selection
\begin{equation}
 F_C = \sum_i m_i(bm_i-cn_i)\,,\,\,\,\, F_D = \sum_i (n_i-m_i)bm_i
\end{equation}
and $N_C = \sum_i m_i$ and $N_D= N - N_C$ are the number of cooperators and defectors respectively. With this notation the change in cooperator frequency \eqref{deltap_s} can be rewritten as
\begin{equation}
 \Delta p(s) = \frac{\delta}{N^2} \left( N_D F_C - N_C F_D \right)  +  {\mathcal O} (\delta^{2})
\end{equation}
This expression was obtained in an intuitive way in Section \ref{overview}. By averaging over the stationary state we of course recover \eqref{dpsel1}.

\subsection{Threshold value from correlations}

Let us now evaluate the expected values in \eqref{critgenexp}. We randomly choose three individuals $k,q$, and $l$ with replacement. All expected values in \eqref{critgenexp} can be expressed in terms of probabilities in the neutral stationary state
\begin{subequations}
\label{exptopr}
\begin{align}
\left\langle \sum_{i} m_{i}^{2} \right\rangle_0 &= N^{2} ~ \mathrm{Pr} (S_k=S_q=1,~ X_k=X_q)
	\label{etpa}\\
\left\langle \sum_{i} m_{i} n_{i} \right\rangle_0 &= N^{2} ~ \mathrm{Pr} (S_k=1 ,~ X_k=X_q)
	\label{etpb}\\
\left\langle \sum_{i,j} m_i m_j n_j \right\rangle_0 &= N^{3} ~ \mathrm{Pr}(S_l=S_k=1,~ X_k=X_q)
	\label{etpc}
\end{align}
\end{subequations}
The indices $i$ and $j$ refer to positions, while $k,q$ and $l$ refer to individuals. These identities are self explanatory, nevertheless they are proven in Appendix \ref{avcorr}.

Because the two strategies are equivalent in the {\it neutral} stationary state, all expressions \eqref{exptopr} remain valid when we change any 1 to 0. Consequently all expressions \eqref{exptopr} simplify to
\begin{equation}
\label{exptopr3}
\begin{split}
\left\langle \sum_{i} m_{i}^{2} \right\rangle_0 &= \frac{N^2}{2} ~ \mathrm{Pr} (S_k=S_q,~ X_k=X_q)\\
\left\langle \sum_{i} m_{i} n_{i} \right\rangle_0 &= \frac{N^2}{2} ~ \mathrm{Pr} (X_k=X_q)\\
\left\langle \sum_{i,j} m_i m_j n_j \right\rangle_0 &= \frac{N^3}{2}~ \mathrm{Pr}(S_l=S_k,~ X_k=X_q)
\end{split}
\end{equation}
Note that these probabilities are denoted in Section \ref{overview} as $P_2$, $P_1$, and $P_3$ respectively. Substituting the probabilities of \eqref{exptopr3} into \eqref{critgenexp} we arrive at the general condition expressed in terms of two and three point correlations
\begin{equation}
\label{critwprob}
	\left(\frac{b}{c} \right)^* = \frac{\mathrm{Pr}(S_l=S_k,~ X_k=X_q) - \mathrm{Pr} (X_k=X_q)}
	{\mathrm{Pr}(S_l=S_k,~ X_k=X_q) - \mathrm{Pr} (S_k=S_q,~ X_k=X_q)}
\end{equation}

In Section \ref{neutral} we have calculated similar probabilities defined in \eqref{corrdefs}, but always for two different individuals. In other words while in the probabilities of \eqref{exptopr3} we pick two individuals with replacement, in the quantities of \eqref{corrdefs} two individuals were picked without replacement. We know, however, that out of two individuals we pick the same individual twice with probability $1/N$, and pick two different individuals otherwise. We also know the corresponding probabilities when picking three individuals. With this  knowledge we can express the probabilities with replacement in \eqref{exptopr3} with the probabilities without replacement in \eqref{corrdefs} as follows
\begin{equation}
\label{exptoprfin}
	\begin{split}
		 \mathrm{Pr} (S_k=S_q,~ X_k=X_q) &= \frac{1}{N} \, \left[ (N-1)g+1\right]\\  
		 \mathrm{Pr} (X_k=X_q) &= \frac{1}{N} \, \left[ (N-1) z+1\right]\\  
 		 \mathrm{Pr}(S_l=S_k,~ X_k=X_q) &= \frac{1}{N^2} \, 
		\left[ (N-1)(N-2) h + (N-1)\left(z+y+g \right) +1 \right]
	\end{split} 
\end{equation}

Now we substitute these probabilities into condition \eqref{critwprob} to obtain the threshold condition
\begin{equation}
\label{critgen}
	\left(\frac{b}{c} \right)^* = \frac{(N-2)(z-h)+ 1- y + z-g }
	{(N-2)(g  - h)+ 1 - y - z+g }
\end{equation}
The above condition \eqref{critgen} is exact for any finite $N$ with self interaction. Without self interaction a ${\mathcal O}(1/N)$ correction appears as discussed in Appendix \ref{exclude}. The model of course makes no sense for $N=1$, and the smallest interesting population size is $N=2$. In the $N\to\infty$ limit of \eqref{critgen} we also obtain a simple rule
\begin{equation}
\label{hami}
 \left(\frac{b}{c} \right)^* = \frac{z-h}{g  - h}
\end{equation}
Substituting the expressions \eqref{dfinal}, \eqref{gfinal}, and \eqref{hfinal} into the above equation for $z$, $g$, and $h$ respectively, we arrive at
\begin{equation}
 \label{bcfinal}
 \left(\frac{b}{c} \right)^* = \frac{\mu C_3 - (2+\mu)C_1+(1+\mu)^2}{\mu C_3 + (2+\mu)C_1-(1+\mu)}
\end{equation}
where we have used the shorthand notation \eqref{shorty}. This is our main result: the exact threshold $b/c$ ratio in the $N\to\infty$ and weak selection limit. For parameter values $b/c>(b/c)^*$ there are more cooperators than defectors in the system in the long time average.

\begin{figure}
\centering
\includegraphics[scale=0.9]{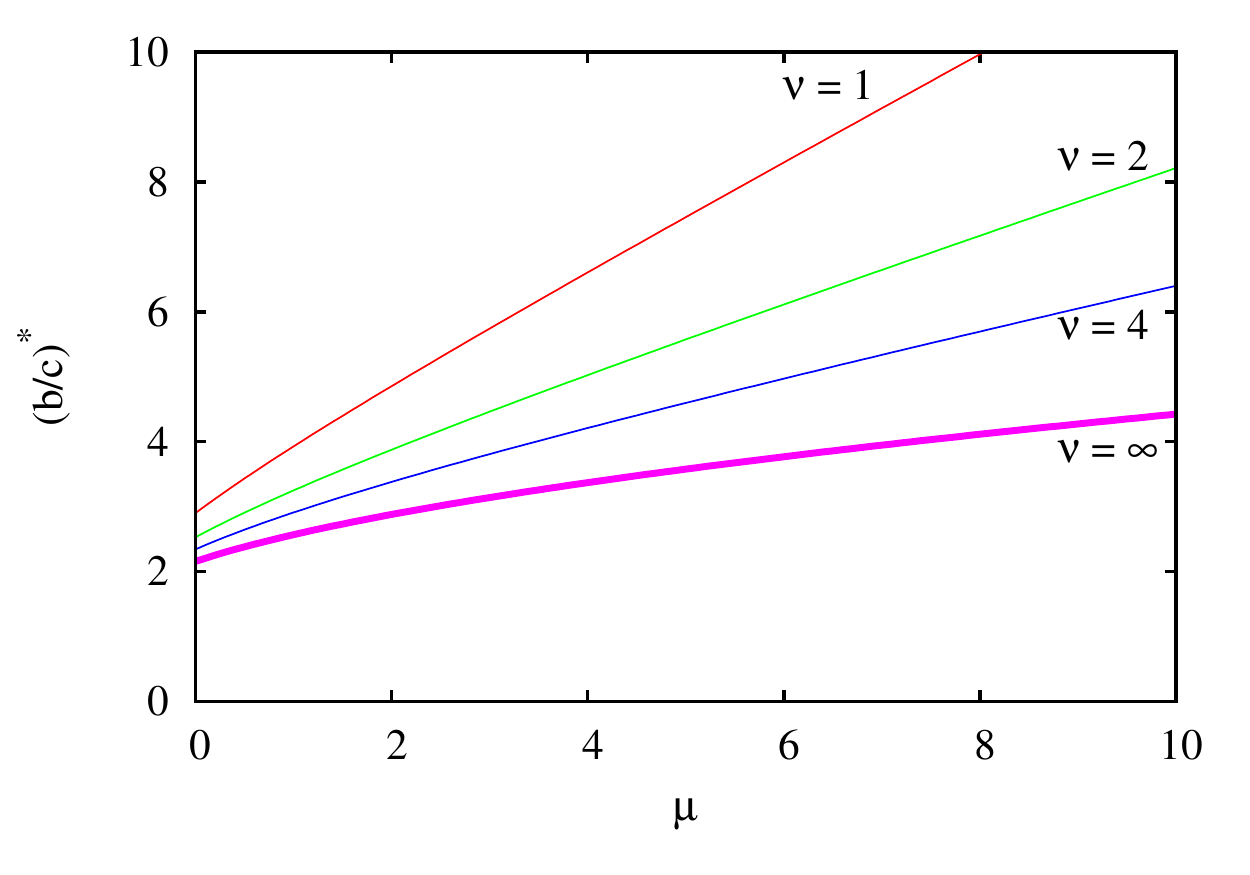}
\caption{Exact threshold $b/c$ ratio \eqref{bcfinal} in the $N\to\infty$ limit for several values of $\nu$. Cooperation is most favored in the $\mu\to 0$ and $\nu\to\infty$ limit, where $(b/c)^*=1+2/\sqrt{3}$.}
\label{bcvermu}
\end{figure}

In Figure~\ref{bcvermu}, we plot the exact $(b/c)^*$ ratio \eqref{bcfinal} as a function of $\mu$ for several values of $\nu$. One observes that $(b/c)^*$ gets smaller both for smaller $\mu$ and for larger $\nu$ . Hence small strategy mutation and large phenotype mutation helps cooperation. The large $\nu$ limit includes the finite $v$ (phenotype changing probability) case. Note that since the cluster size in phenotype space is $\sqrt{2Nv}$, the average number of individuals with the same phenotype is proportional to $\sqrt{N/v}$, hence there are plenty of individuals to interact with even for finite $v$ values in the large $N$ limit.

In the $\nu\to\infty$ limit \eqref{bcfinal} becomes
\begin{equation}
\label{bclarger}
	\left(\frac{b}{c} \right)^* = \frac{ (2+\mu)\sqrt{1+\mu} -2(1+\mu)^2 -\mu\sqrt{3+\mu}}
	{ - (2+\mu)\sqrt{1+\mu} +2(1+\mu)- \mu\sqrt{3+\mu}}
	+ {\mathcal O}(\frac{1}{\sqrt \nu})
\end{equation}
which for $\mu\to 0$ behaves as
\begin{equation}
 \label{bcbigrsmallmu}
 \left(\frac{b}{c} \right)^* = 1+ \frac{2\sqrt{3}}{3} + \mu \frac{7\sqrt{3}-3}{18} + {\mathcal O}(\mu^2)
\end{equation}
which is $\approx 2.16$ in the leading order. For $\mu\to\infty$ the threshold ratio \eqref{bclarger} diverges as
\begin{equation}
	\left(\frac{b}{c} \right)^* = \sqrt{\mu}  + 1 + {\mathcal O}(\frac{1}{\sqrt\mu})
\end{equation}

Conversely, in the $\mu\to0$ limit \eqref{bcfinal} becomes
\begin{equation}
\label{bcsmallmu}
	\left(\frac{b}{c} \right)^* = \frac{\sqrt{3}(1+4\nu)^{3/2}+(3+8\nu)\sqrt{3+4\nu}}{\sqrt{3}(1+4\nu)^{3/2}-\sqrt{3+4\nu}}
	+ {\mathcal O}(\mu)
\end{equation}
This limit function diverges as $3/4\nu$ for small $\nu$, but converges to the constant $1+2/\sqrt{3}$ as $\nu\to\infty$. Hence the best scenario for cooperation is $\mu\to 0$ and $\nu\to\infty$ where $(b/c)^*=1+2/\sqrt{3}$.

The large $N$ asymptotic results are identical for the Moran process, where we choose a random individual to die, and another (with replacement) to reproduce with probability proportional to the player's payoff (see Appendix \ref{Moran}).

We would like to briefly comment on the relationship between our work and inclusive fitness or kin selection theory \cite{hamilton64,rousset03,rousset07}. Let $R$ be the inverse of the r.h.s.\ of \eqref{critwprob}. Now we formally obtained Hamilton's rule $(b/c)^*=1/R$. By dividing both the numerator and the denominator in $R$ by $\mathrm{Pr} (X_k=X_q)$, (we can assume that it is not zero), and using the definition of conditional probability, we can rewrite $R$ as
\begin{equation}
\label{relatcond}
	R = \frac
	{\mathrm{Pr} (S_k=S_q|~ X_k=X_q) - \mathrm{Pr}(S_l=S_k|~ X_k=X_q)}
	{1 - \mathrm{Pr}(S_l=S_k|~ X_k=X_q) }
\end{equation}
Now with the notation
\begin{equation}
\label{defs}
 G = \mathrm{Pr} (S_k=S_q|~ X_k=X_q), \quad \overline G = \mathrm{Pr}(S_l=S_k|~ X_k=X_q)
\end{equation}
we obtain $R = (G-\overline G)/(1 - \overline G)$, which is in the form of usual relatedness formula. Note, however, that this $\overline G$ is not the probability of identity in state (IIS) between two random individuals in the population as it usually is in inclusive fitness theory. Instead, $\overline G$ is a sort of weighted average of IIS probabilities in which those who share the same phenotype with more players are assigned a larger weight.

\section{Conclusions}

We have derived the conditions for cooperation to be favored for games in phenotype space for any population size and mutation rate. Figure \ref{fig:fit} shows the excellent agreement between numerical simulations and analytical calculations. The argument that leads to inequality \eqref{avercond} contains self-interaction, which means that each cooperator adds $b-c$ to his payoff. Typically, self-interaction is not a desirable assumption, but it does simplify the calculation. Excluding self-interaction requires us to calculate two more correlation terms (see Appendix~\ref{exclude}). But in the limit of large population size, the difference between the two approaches results only in a $1/N$ correction term for the critical benefit-to-cost ratio. Thus, the crucial condition \eqref{condi} holds for the case with and without self-interaction. 

In Appendix \ref{general} we expanded our analysis to study any $2\times 2$ game, not only the interaction between cooperators and defectors. Here the general payoff matrix is given by \eqref{payoffmatgen}. For the game in a one dimensional phenotype space and large population size we find that $C$ is more abundant than $D$ if 
\begin{equation}
\label{gencondi}
(R-P)(1+\sqrt{3}) > T-S.
\end{equation}
This formula can be used for evaluating any two-strategy symmetric game in a one dimensional phenotype space. We have discussed the snow-drift game and the stag-hunt game as particular examples.

We can also study higher dimensional phenotype spaces. In general, for higher dimensions it is easier for cooperators to overcome defectors. The intuitive reason is that in higher dimensions phenotypic identity also implies strategic identity. In Appendix \ref{infdim}, we show that in the limit of infinitely many dimensions, and under the same assumptions that produced conditions \eqref{condi} and \eqref{gencondi}, 
the crucial benefit-to-cost ratio in the Prisoner's Dilemma converges to $b/c>1$. For general games, the equivalent result of condition \eqref{gencondi} becomes $R>P$, which means the evolutionary process always chooses the strategy with the higher payoff against itself.  Our basic approach can also be adapted to continuous, rather than discrete, phenotype spaces. In this case, no two individuals have exactly the same phenotype, but the conditional behavioral strategy is triggered by sufficient phenotypic similarity.
 
In summary, we have developed a model for the evolution of cooperation based on phenotypic similarity. Our approach builds on previous ideas of tag based cooperation, but in contrast to earlier work  \cite{{axelrod04},{jansen06},{hammond06},{rousset07},{gardner07}}, we do not need spatial population dynamics to obtain an advantage for cooperators. We derive a completely analytic theory that provides general insights.
We find that the abundance of cooperators in the mutation-selection equilibrium is an increasing function of the phenotypic mutation rate and a decreasing function of the strategic mutation rate. These observations agree with the basic intuition that higher phenotypic mutation rates reduce the interactions between cooperators and defectors, while higher strategic mutation rates destabilize clusters of cooperators by allowing frequent invasion of newly mutated defectors. Therefore, cooperation is more likely to evolve if the strategy mutation rate is small and if the phenotypic mutation rate is large. In a genetic model this assumption may be fulfilled if the strategy is encoded by one or a few genes, while the phenotype is encoded by many genes. Also in a cultural model, it can be the case that the phenotypic mutation rates are higher than the strategic mutation rates: for example, people might find it easier to modify their superficial appearance than their fundamental behaviors. Furthermore, we show how the correlations between strategies and phenotypes can be obtained from neutral coalescence theory under the assumption that selection is weak \cite{{wakeley08},{rousset03}}. Our theory can be applied to study any evolutionary game in the context of conditional behavior that is based on phenotypic similarity or difference. 

\begin{acknowledgments}
We are grateful for support from the John Templeton Foundation, the NSF/NIH (R01GM078986) joint program in mathematical biology, the Bill and Melinda Gates Foundation (Grand Challenges grant 37874), the Japan Society for the Promotion of Science, and J. Epstein.
\end{acknowledgments}

\appendix


\section{Finite populations for $v=1/2$}
\label{wf}

Here we consider the Wright-Fisher (W-F) model for finite $N$ and $v=1/2$. What makes this case simple is that at each time step all individuals move. The probability that the time to the MRCA is $t$ is given by \eqref{timewf}. During $t$ generations there are exactly $2t$ birth events in the ancestry of two individuals, and in the $v=1/2$ case the phenotypic distance between two individuals follows a simple random walk with two steps in phenotype space per one time unit. Consequently, the distance between two siblings is always even. After some transient time the whole population will be constrained on the same sub-lattice of even, and then odd sites. The distance distribution of two individuals $k$ and $q$, time $t$ after their MRCA is
\begin{equation}
\label{spacewf}
 \mathrm{Pr}(X_k-X_q=x|T=t) = 2^{-2t} \binom{2t}{t+x/2}
\end{equation}
where again $x$ is always even. Consequently the probability $z(x)$ that two randomly chosen individuals are at distance $x$ apart can be obtained from \eqref{dismethod}
\begin{equation}
 z(x) = \frac{1}{N-1} \sum_{t=1}^\infty \binom{2t}{t+x/2} \left( \frac{N-1}{4N}\right)^{t}
\end{equation}
This sum can be evaluated using the identity
\begin{equation}
\label{sumrule}
 \sum_{t=1}^\infty \binom{2t}{t+x/2} \left( \frac{a}{4}\right)^t = 
 \begin{cases}
 \frac{a}{\sqrt{1-a}(1+\sqrt{1-a})}, & x=0\\
 \frac{a^{|x|/2}}{\sqrt{1-a}(1+\sqrt{1-a})^{|x|}}, & |x|\ge 2
 \end{cases}
\end{equation}
to obtain
\begin{equation}
\label{distwf}
 z(x) = 
 \begin{cases}
 \displaystyle \frac{1}{\sqrt N+1} & x=0\\
 \displaystyle \frac{\sqrt N}{N-1} \left( \frac{N-1}{N+2\sqrt N +1} \right)^{|x|/2} & |x|\ge 2
 \end{cases}
\end{equation}
Hence, apart from the special $x=0$ case, $z(x)$ decays exponentially in $x$. For fixed distances and $N\to\infty$ the asymptotic behavior is $z(x)=1/\sqrt N +{\mathcal O}(1/N)$. The second moment of the distance distribution \eqref{distwf} is simply $2N$.

Now we turn to the strategies of the individuals. The strategies of the two players are the same if no mutations happened during time $t$ to either player, which is the case with probability $(1-u)^{2t}$. Otherwise the two strategies are the same with probability $1/2$. Consequently, the conditional probability is
\begin{equation}
\label{evenwf}
 y(t) = (1-u)^{2t} + \frac{1}{2}[1-(1-u)^{2t}] = \frac{1+U^t}{2}
\end{equation}
where we introduce the shorthand notations
\begin{equation}
 U = (1-u)^2\,\,,\,\,\,\,\, M = N(1-U)+U
\end{equation}
The probability $y$ that two randomly chosen individuals have the same strategy becomes
\begin{equation}
\label{sigmawf}
 y = \sum_{t=1}^\infty p(t) y (t) = \frac{1}{2} \left( 1+\frac{U}{M} \right)
\end{equation}
where we have used \eqref{timewf} and \eqref{evenwf}.

Similarly, using \eqref{gmethod} we obtain the probability $g$ that two randomly chosen individuals have both the same strategy and the same phenotype
\begin{equation}
\label{wfcc0}
 g = \frac{1}{2(\sqrt{N} + 1)} + \frac{U}{2\sqrt{M}\left(\sqrt{N} + \sqrt{M}\right)} 
\end{equation}
These are exact results for arbitrary number of individuals $N$ and mutation rate $u$. 
In the $N\to\infty$ and $u\to0$ limit of the formulas \eqref{distwf}, \eqref{sigmawf} and \eqref{wfcc0} with $\mu=2Nu$ kept constant, we recover the $\nu\to\infty$ limits of the corresponding formulas \eqref{distance}, \eqref{sigma} and \eqref{gfinal}, apart from a factor two. This factor two is a peculiarity of the $v=1/2$ case. Since here the distance between individuals is always even, there must be twice as many players at a given even distance. Note also that the variance of the cluster is $2\nu$ both for $v=1/2$ and for the continuous limit calculation.

For only two individuals, the general condition \eqref{critgen} simplifies to
\begin{equation}
\label{critgenN=2}
	\left(\frac{b}{c} \right)^*_{N=2} = \frac{ 1-y + z -g }{1-y - z+g }
\end{equation}
which contains only quantities we have just calculated in this section. To obtain the exact $(b/c)^*$ for any other finite $N$ we have to use the general expression \eqref{critgen}, and obtain $h$ analogously to \eqref{threeform} and using \eqref{tt1} and \eqref{tt2}. The formulas for $h$ and $(b/c)^*$ are too cumbersome to include here. We have, however, checked these formulas with computer simulations for many values of $N$. We explicitly simulated the W-F process and found the threshold $(b/c)^*$ value where the frequency of cooperators in the stationary state becomes larger than $1/2$. Moreover, in the $N\to\infty$, $u\to 0$ limit with $\mu=2Nu$ constant, we recover the continuous time formula \eqref{bclarger}.

\section{Excluding self interaction}
\label{exclude}

If cooperators cannot interact with themselves, we have
\begin{equation}
	\begin{split}
		f_{C, i} &= 1+ \delta \left[ b (m_{i}-1) - c (n_{i}-1) \right]\\
		f_{D, i} &= 1+ \delta \left[ b m_{i} \right].
	\end{split}
\end{equation}
Therefore the fitness of cooperators at position $i$ becomes
\begin{equation}
	\begin{split}
		w_{C, i} &= 1 + \frac{\delta}{N} \left( b (m_{i}-1) - c (n_{i}-1) - \frac{b-c}{N} \sum_{j} m_{j} (n_{j}-1) \right) +  {\mathcal O} (\delta^{2})\\
	\end{split}
\end{equation}
which then leads to the expected change of cooperator frequency
\begin{equation}
	\begin{split}
	\left\langle \Delta p \right\rangle =& \frac{\delta}{N^2} \Bigg[ b \left\langle \sum_{i} m_{i}^{2} \right\rangle - c  \left\langle \sum_{i} m_{i} n_{i} \right\rangle - \frac{b-c}{N} \left\langle \sum_{i,j} m_{i} m_{j} n_{j} \right\rangle\\
	& -(b-c) \left\langle \sum_{i} m_{i} \right\rangle  + \frac{b-c}{N} \left\langle \sum_{i,j} m_{i} m_{j} \right\rangle \Bigg] +  {\mathcal O} (\delta^{2}).
	\end{split}
\end{equation}
Two new correlation types in the neutral stationary state appear 
\begin{equation}
\begin{split}
\left\langle \sum_{i} m_{i} \right\rangle &= N ~ \mathrm{Pr} (S_k=1) = \frac{N}{2}\\
\left\langle \sum_{i,j} m_{i} m_{j} \right\rangle &= N^{2} ~ \mathrm{Pr} (S_k=S_q=1)
  = \frac{N^2}{2} y \\
\end{split}
\end{equation}
This then leads to the general expression analogous to \eqref{critgen} for the threshold ratio
\begin{equation}
\label{critgennoself}
	\left(\frac{b}{c} \right)^* = \frac{(N-2)(z-h)+  z-g }
	{(N-2)(g  - h) - z+g }
\end{equation}
The smallest valid population size is $N=3$. In the $N\to\infty$ the threshold $b/c$ ratio with self interaction \eqref{critgen} and without it \eqref{critgennoself} are the same \eqref{hami} in the leading order, and their difference is only of order $1/N$.

\section{From averages to correlations}
\label{avcorr}

Here we obtain the identities listed in \eqref{exptopr}. The variables $m_i$ and $n_i$ are fixed in any given state. Let us use the indicator function $\mathbbm{1}$, which is $\mathbbm{1}(A)=1$ if event $A$ is true and $\mathbbm{1}(A)=0$ if event $A$ is false. Of course the stationary average of the indicator function is the stationary probability of an event 
\begin{equation}\label{indprob}
\langle \mathbbm{1}(A) \rangle = \mathrm{Pr} (A)
\end{equation}
and by $\mathbbm{1}(A, B)$ we mean $\mathbbm{1}(A ~\cap~ B)=\mathbbm{1}(A)\mathbbm{1}(B)$. Now in any given state we can express $n_i$ and $m_i$ by the indicator functions
\begin{equation}
	\begin{split}
		 n_i &= \sum_k \mathbbm{1}(X_k=i)\\
		 m_i &= \sum_q \mathbbm{1}(X_q=i) \mathbbm{1}(S_q=1).
	\end{split} 
\end{equation}

The sum in \eqref{etpa} becomes
\begin{equation}\label{suma1}
 	\sum_i m_i m_i = \sum_{k,q} \left[  \mathbbm{1}(S_k=1) \mathbbm{1}(S_q=1) 
	\sum_i \mathbbm{1}(X_k=i)  \mathbbm{1}(X_q=i)\right] = \sum_{k,q} 
	\mathbbm{1}(S_k=S_q=1) \mathbbm{1}(X_k=X_q)
\end{equation}
since the sum over $i$ is simply
\begin{equation}
	\sum_i \mathbbm{1}(X_k=i)  \mathbbm{1}(X_q=i) = \sum_i \mathbbm{1}(X_k=i ,~ X_q=i) = 
 \mathbbm{1}(X_k=X_q).
\end{equation}
Now taking the average of \eqref{suma1} in the stationary state we obtain
\begin{equation}
  \left\langle \sum_i m_i^2 \right\rangle_0 
  = \sum_{k,q} \langle \mathbbm{1}(S_k=S_q=1 ,~ X_k=X_q) \rangle
  = \sum_{k,q} \mathrm{Pr} (S_k=S_q=1 ,~ X_k=X_q),
\end{equation}
where we have used identity \eqref{indprob}. Since all individuals are equivalent in the stationary state, the above probabilities are the same for any pair of individuals, hence from now on we consider $k$ and $q$ as two randomly chosen individuals, and write
\begin{equation}\label{withprobs}
	\left\langle \sum_i m_i^2 \right\rangle_0 = N^2\, \mathrm{Pr} (S_k=S_q=1 ,~ X_k=X_q).
\end{equation}

The expression \eqref{etpb} can be derived similarly, since
\begin{equation}\label{sum1}
 	\sum_i m_i n_i = \sum_{k,q} \left[  \mathbbm{1}(S_q=1) \sum_i \mathbbm{1}(X_k=i)  \mathbbm{1}(X_q=i)\right] = \sum_{k,q} \mathbbm{1}(S_q=1) \mathbbm{1}(X_k=X_q)
\end{equation}
and taking the average of \eqref{sum1} in the stationary state leads to
\begin{equation}
  \left\langle \sum_i m_i n_i \right\rangle_0 
  =  \sum_{k,q} \mathrm{Pr} (S_q=1 , X_k=X_q)
  =  N^2~ \mathrm{Pr} (S_q=1 , X_k=X_q)
\end{equation}

For the last expression \eqref{etpc} we have
\begin{equation}
\begin{split}
 \sum_{i,j} m_i m_j n_j &=  \sum_{k,q,l} \left[ \sum_i \mathbbm{1}(S_l=1, X_l=i) \right]
 \left[ \sum_j \mathbbm{1}(S_k=1, X_k=j) \mathbbm{1}(X_q=j) \right]\\
 &= \sum_{k,q,l} \mathbbm{1}(S_l=1) ~ \mathbbm{1}(S_k=1,~ X_k=X_q)
 \end{split}
\end{equation}
which in the stationary state becomes
\begin{equation}
  \left\langle \sum_{i,j} m_i m_j n_j  \right\rangle_0 
  = \sum_{k,q,l} \mathrm{Pr} (S_l=S_k=1,~ X_k=X_q)
  = N^3~  \mathrm{Pr} (S_l=S_k=1,~ X_k=X_q)
\end{equation}


\section{Moran dynamics} 
\label{Moran}

In the Moran model we chose a random individual to die, and another (with replacement) to multiply with probability proportional to the player's payoff. The newborn then replaces the dead individual. Otherwise the dynamics is the same as in the W-F case. The behavior of the Moran model is also very similar to the W-F model, and the results can be written in an identical form in the $N\to\infty$ limit, by defining the appropriate variables.

We consider the neutral case of the Moran model first. Let us obtain the probability $\mathrm{Pr}(T=t)$ that the time to the most recent common ancestor (MRCA) of two randomly chosen individual is $T=t$. Let us calculate the probability $P_\mathrm{CA}$ that they had a common ancestor one update step before. It could happen only if the parent and the dying individuals were different, which happens with probability $1-1/N$. Then our two individuals have a common ancestor if one of them is the parent and the other is the newborn daughter, which has a probability $2~ \frac{1}{N}~\frac{1}{N-1}$. Hence having a common ancestor in the previous update step is
\begin{equation}
 P_\mathrm{CA} = \left( 1-\frac{1}{N}\right)\cdot 2\cdot \frac{1}{N}\cdot \frac{1}{N-1} = \frac{2}{N^2}
\end{equation}
Consequently the probability that the MRCA is exactly time $T=t$ backward is
\begin{equation}
\label{time}
 \mathrm{Pr}(T=t) = (1-P_\mathrm{CA} )^{t-1} P_\mathrm{CA}  = \left( 1-\frac{2}{N^2} \right)^{t-1} \frac{2}{N^2}
\end{equation}
If we introduce a rescaled time $\tau=t/(N^2/2)$, then in the $N\to\infty$ limit the coalescent time distribution \eqref{time} converges to the same density function \eqref{coaltime} as we obtained for the W-F model.

Since in our model mutations (in strategies) and motion only happen at birth events, let us investigate the statistics of birth events in the Moran model. As we follow the ancestral lines of two randomly chosen individuals backward in time, we can obtain the probability $P_\mathrm{B}$ that a birth event happens in one update step, but the ancestral lines do not coalesce. In other words, $P_\mathrm{B}$ is the probability that at a given time one of the two individuals is the daughter but the other is not the parent. If the parent dies during this update step (which happens with probability $1/N$) one individual is the daughter with probability $2/N$ (and the other individual cannot be the parent). If the parent does not die (which happens with probability $1-1/N$) one of the individuals is the daughter and the other is not the parent with probability $2/N\cdot(N-2)/(N-1)$. Hence the probability that there is a birth event in the ancestry of either individual during one elementary time step is
\begin{equation}
 P_\mathrm{B} 
 = \left( 1-\frac{1}{N}\right)\cdot \frac{2}{N} \cdot \frac{N-2}{N-1} + \frac{1}{N}\cdot \frac{2}{N}
 = \frac{2(N-1)}{N^2} 
\end{equation}
In the continuous time limit with $\tau=t/(N^2/2)$, a birth event happens at rate $N$. Consequently a mutation happens at rate $\mu=Nu$ on the ancestral line of {\it two} individuals. Similarly, one of the two individual hops at rate $\nu=Nv$ in each direction. In other words the distance between the two individuals changes at rate $\nu$ in each direction. This means that the continuous time ($N\to\infty$) descriptions of the Moran and the W-F models are the same, but $N$ must be used for the Moran and $2N$ for the W-F model in the definition of $\mu$ and $\nu$. Hence all $N\to\infty$ results of Section~\ref{neutral} are also valid for the Moran model. (Note that the diffusion coefficient of the cluster is $D=v/N$.)

All formulas of Section~\ref{critical} are almost identical to those for W-F model. The average frequency of cooperators depends on the change of cooperators very similarly to \eqref{averwithsel}
\begin{equation}
 \langle p\rangle = \frac{1}{2} + N\frac{1-u}{u} \langle \Delta p \rangle_\mathrm{sel}
\end{equation}
Instead of the fitness of the W-F model \eqref{fittWF}, we have a very similar expression for the fitness after one elementary step
\begin{equation}
\label{fittM}
	\begin{split}
		w_{C, i} &= \frac{N-1}{N} + \frac{f_{C,i}}{\sum_{j} [ m_{j} f_{C,j} + (n_{j}-m_{j}) f_{D,j}]}\\
	\end{split}
\end{equation}
where the payoffs are again given by \eqref{payoff}.
Here the first term corresponds to the cooperator staying alive, and to second to it being chosen for reproduction. In the $\delta\to 0$ limit \eqref{fittM} becomes
\begin{equation}
\label{fittfinalM}
		w_{C, i} = 1 + \frac{\delta}{N} \left( b m_{i} - c n_{i} - \frac{b-c}{N} \sum_{j} m_{j} n_{j} \right) +  {\mathcal O} (\delta^{2})\\
\end{equation}
Note that this is exactly the fitness of the W-F process \eqref{fittfinalWF} with a scaled selection strength $\delta'=\delta/N$. Hence all results of Section~\ref{critical}, and in particular the citical $b/c$ ratio \eqref{bcfinal} are also valid for the Moran model.

\section{General payoff matrix}
\label{general}

Instead of the payoff matrix \eqref{payoffmat} of the simplified Prisoner's Dilemma (PD) game, we study now a general payoff matrix
\begin{equation}
  \label{payoffmatgen}
  \begin{pmatrix} R & S \\ T & P \end{pmatrix}
\end{equation} 
A similar derivation to the one presented in Section~\ref{critical} leads to the condition for cooperation
\begin{equation}
\label{ggamecond}
 (R-S)g+(S-P)z>(R-S-T+P)\eta+(S+T-2P)h
\end{equation}
in the $N\to\infty$ limit, which is the analogous formula to \eqref{hami}. Here a new type of three point correlation must be introduced
\begin{equation}
\label{etadef}
 \eta = \mathrm{Pr} (S_l=S_k=S_q,~ X_k=X_q )
\end{equation}
In the $\nu\to\infty$ and $\mu\to 0$ limit the correlations are
\begin{equation}
 \label{corrlimit}
 \begin{tabular}{ll}
  $\displaystyle z = \frac{1}{2\sqrt \nu}$&
  $\displaystyle g =  \frac{1}{2\sqrt \nu}\left( 1 - \frac{\mu}{4} \right)$\\
  $\displaystyle h =  \frac{1}{2\sqrt \nu}\left( 1 - \mu \frac{2+\sqrt 3}{8} \right)$\,\,\,\,\,\,\, &
  $\displaystyle \eta =  \frac{1}{2\sqrt \nu}\left( 1 - \mu \frac{3+\sqrt 3}{8} \right)$\\
 \end{tabular}
\end{equation}
up to ${\mathcal O}(1/\nu)$ and ${\mathcal O}(\mu^2)$ terms. Here $z$, $g$, and $h$ were obtained as limits of the general expressions \eqref{dfinal}, \eqref{gfinal}, and \eqref{hfinal} respectively. The value of $\eta$ was derived analogously to \eqref{threeform}. By substituting these correlations into \eqref{ggamecond} we finally arrive at the general condition for cooperation 
\begin{equation}
\label{cute}
 T-S < (R-P)(1+\sqrt{3})
\end{equation}
For the simplified PD game \eqref{payoffmat} we recover \eqref{bcbigrsmallmu} in the leading order.

\begin{figure}
\centering
\includegraphics[scale=0.9]{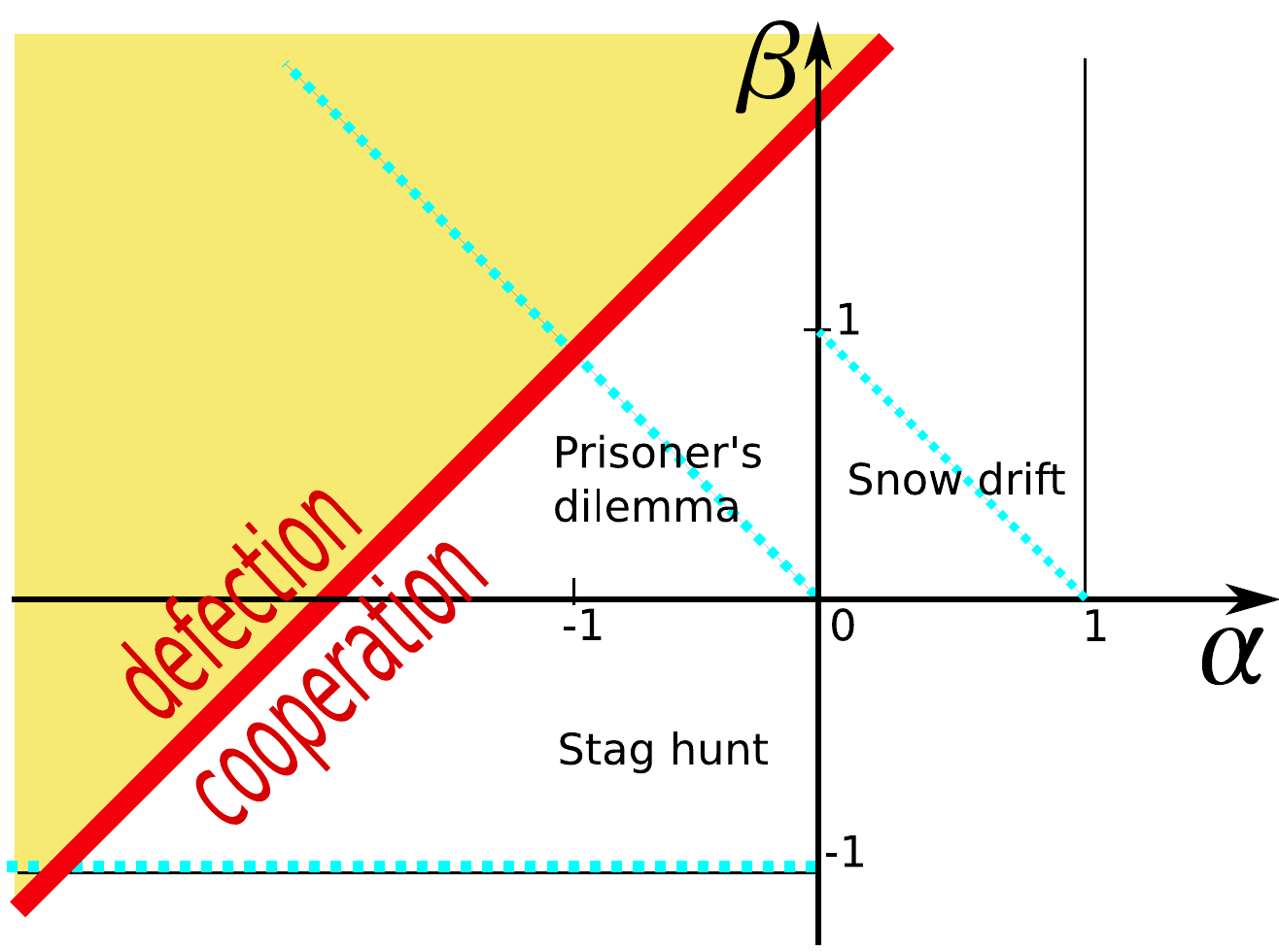}
\caption{``Snow drift", ``Stag hunt" and ``Prisoner's dilemma" games correspond to three distinct regions in the $(\alpha, \beta)$ plane, bounded by black lines. The red (thick) line \eqref{straight} marks the boundary between defection (yellow-shaded) and cooperation (white). The blue (thicker dashed) lines depict the corresponding simplified payoff matrices.}
\label{st}
\end{figure}

For a non-degenerate payoff matrix, with the exchange of players $R>P$ can always be achieved. Then under weak selection one can define an equivalent matrix
\begin{equation}
  \label{payoffmatgeneq}
  \begin{pmatrix} 1 & \alpha \\ 1+\beta & 0 \end{pmatrix}
\end{equation} 
with only two parameters
\begin{equation}
 \alpha = \frac{ S- P}{ R- P} \, , \,\,\,\,\, \beta = \frac{T- R}{ R- P}
\end{equation}
In these variables the condition for cooperation \eqref{cute} becomes
\begin{equation}
\label{straight}
 \beta < \alpha +\sqrt{3}
\end{equation}
which describes a straight threshold line in the $(\alpha, \beta)$ plane (see Figure~\ref{st}). 

In Figure~\ref{st} we show how this threshold line \eqref{straight} divides the $(\alpha, \beta)$ plane into a cooperative and a defective half plane. Three regions, bounded by black lines, correspond to the ``Snow drift", the ``Stag hunt" and the ``Prisoner's dilemma" games. The blue straight lines on the $(\alpha, \beta)$ plane correspond to the following representative simplified payoff matrixes
\begin{equation}
\begin{tabular}{lll}
 Snow drift & $\begin{pmatrix} b-c/2 & b-c\\ b & 0 \end{pmatrix}$  \,\,\,\, 
 	& $\beta = 1-\alpha$, with $0<\alpha<1$\\
 Stag hunt & $\begin{pmatrix} b-c & -c\\ 0 & 0 \end{pmatrix}$ 
 	& $\beta = -1$, with $\alpha<0$\\
 Prisoner's dilemma & $\begin{pmatrix} b-c & -c\\ b & 0 \end{pmatrix}$ 
 	& $\beta = -\alpha$, with $\alpha<0$\\
\end{tabular} 
\end{equation}
Form the general condition \eqref{cute} we can deduce the condition for cooperation for these simplified games. There is always cooperation in the simplified Snow drift game. Cooperation is favored in the simplified Stag hunt game only for $b/c>1+1/(1+\sqrt 3)$. In the simplified PD game cooperators win for $b/c>1+2/\sqrt 3$ in agreement with \eqref{bcbigrsmallmu}.


\section{Randomly changing phenotypes}
\label{infdim}

Here we replace the one-dimensional phenotype space with an infinite-dimensional phenotype space.  We do not model the number of dimensions explicitly, but simply assume that every mutation causes a jump to a new unique phenotype.  Now the only way that two individuals can have the same phenotype is if there are no phenotypic mutations in their ancestry back to the time of their most recent common ancestor.  This property is called \textit{identity by descent} in population genetics and this mutation model known as the infinitely-many-alleles, or simply infinite-alleles, mutation model \cite{malecot46,kimura64}.

Let $\tilde v$ be the probability that the phenotype of an offspring differs from that of its parent. Note that in the one-dimensional model, there is a mutation probability of $v$ in each direction.  As before, in the limiting ($N\to\infty$) model with time rescaled appropriately, the phenotypic mutation rate to two individuals is equal to $\nu$.  In the Wright-Fisher model we have $2N\tilde v\to \nu$ (and $N\tilde v\to \nu$ in the Moran model), where the arrows correspond to the limit $N\to\infty$. The definition of $\mu=2Nu$ in the  Wright-Fisher model ($\mu=Nu$ in the Moran model) is the same as before.

\begin{figure}
\centering
\includegraphics[scale=0.9]{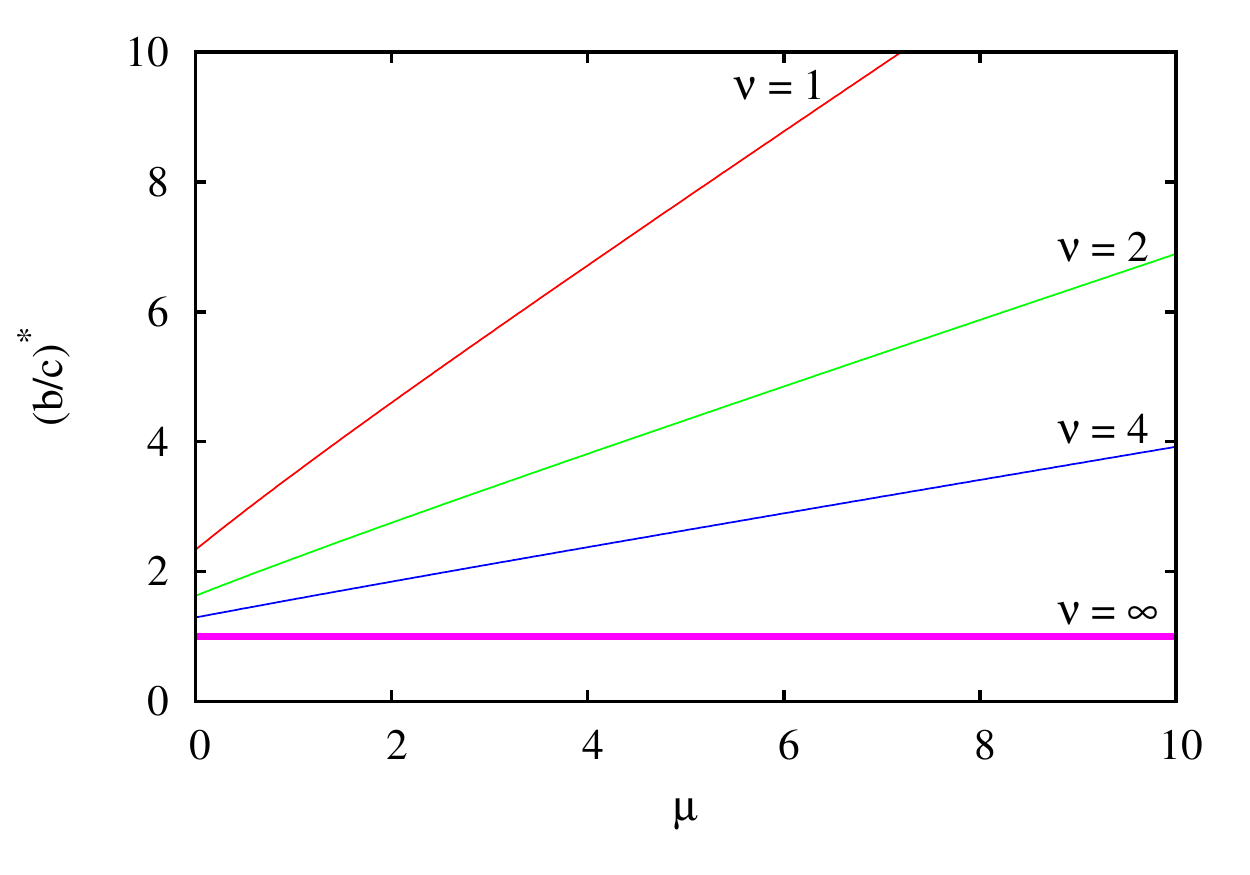}
\caption{Exact threshold $b/c$ ratio (\ref{eq:bcinf}) for randomly changing phenotypes for $N\to\infty$. Cooperation is most favored in the $\nu\to\infty$ limit, where $(b/c)^*=1$. Note that the lines for finite values of $\nu$ are not straight.}
\label{bcinfdim}
\end{figure}

Given a coalescence time $\tau$ between a pair of individuals, 
\begin{equation}
\zeta(\tau) = e^{-\nu\tau} \label{eq:ztau}
\end{equation}
is the probability that they have the same phenotype.  Therefore, in the $N\to\infty$ limit, the correlations defined in \eqref{corrdefs} become
\begin{equation}
\label{infcorr}
\begin{split}
z &=  \frac{1}{1+\nu} \\
g &= \frac{1}{2} \left( \frac{1}{1+\nu} + \frac{1}{1+\mu+\nu}\right)\\
h &=  \frac{1}{2} \left[ \frac{1}{1+\nu} + 
	\frac{1}{3+\mu+\nu} \left( \frac{1}{1+\nu} + \frac{1}{1+\mu}+\frac{1}{1+\mu+\nu} \right) \right]
\end{split}
\end{equation}
The calculation goes analogously to that of Section \ref{neutral}.
The threshold parameters \eqref{hami} for cooperation to be favored becomes  
\begin{equation}
\label{eq:bcinf}
\left(\frac{b}{c}\right)^{\ast} = \frac{\nu(3+2\mu+\nu) + (1+\mu)(3+\mu)}{\nu(2+\mu+\nu)} 
\end{equation}
This is plotted in Figure~\ref{bcinfdim}, which can be compared to the corresponding Figure~\ref{bcvermu} for the one-dimensional model.  

Cooperation is most favored when $\nu$ is large because in this case two individuals that share the same phenotype will almost surely have the same strategy.  We have 
\begin{equation}
\label{eq:bcinflarger}
\left(\frac{b}{c}\right)^{\ast} = 1 + \frac{1+\mu}{\nu} + O(\nu^{-2}) 
\end{equation} 
In the $\nu\to\infty$ limit, $(b/c)^*=1$, {\it i.e.} cooperation is favored whenever the benefit $b$ from cooperation is larger than the cost $c$.

For general payoff matrices \eqref{payoffmatgen}, we restrict our calculation to the $\mu\to0$ limit. The calculation is completely analogous to that of Appendix \ref{general}. First we calculate the three point correlation $\eta$, which is defined in \eqref{etadef}. Up to first order in $\mu$ we obtain
\begin{equation}
 \eta = \frac{1}{1+\nu} \left[ 1-\mu\frac{9+7\nu+2\nu^2}{4(1+\nu)(3+\nu)} \right]
\end{equation}
Substituting this expression together with \eqref{infcorr} into the general condition \eqref{ggamecond} for cooperation, we finally obtain
\begin{equation}
\label{inffinal}
 T-S < (R-P)\, \frac{(1+\nu)(3+2\nu)}{3+\nu}
\end{equation}
This result is valid for general values of $\nu$. For $\nu\to0$ condition \eqref{inffinal} becomes $T-S<R-P$, while in the $\nu\to\infty$ limit it is simply $R>P$.

By using the scaled variables $\alpha$, $\beta$, introduced in \eqref{payoffmatgeneq}, condition \eqref{inffinal} is again a straight line in the $(\alpha, \beta)$ plane. For $\nu\to0$ there is no cooperation in the PD region (see this region in Figure \ref{st}), but for $\nu\to\infty$ the whole plane corresponds to cooperation.

\end{document}